\documentclass[a4paper,11pt]{article}
\usepackage{pos}

\title{Shedding new light on the Hubble constant tension through Supernovae Ia}
\ShortTitle{SNe Ia and the Hubble tension}

\author*[a,b,c,d,e]{Maria Giovanna Dainotti}
\author[f,g]{Biagio De Simone}
\author[h,i]{Giovanni Montani}
\author[j,k]{Malgorzata Bogdan}

\affiliation[a]{Division of Science, National Astronomical Observatory of Japan, 2-21-1 Osawa, Mitaka, Tokyo 181-8588, Japan}
\affiliation[b]{The Graduate University for Advanced Studies (SOKENDAI), Shonankokusaimura, Hayama, Miura District, Kanagawa 240-0115}
\affiliation[c]{Space Science Institute, 4765 Walnut St Ste B, Boulder, CO 80301, USA}
\affiliation[d]{Nevada Center for Astrophysics, University of Nevada, 4505 Maryland Parkway, Las Vegas, NV 89154, USA}
\affiliation[e]{Bay Environmental Institute, P.O. Box 25 Moffett Field, CA, California}
\affiliation[f]{Dipartimento di Fisica, Universit\'a di Salerno, Via Giovanni Paolo II, 132 I-84084 Fisciano (SA), Italy}
\affiliation[g]{INFN, Sezione di Napoli, Gruppo collegato di Salerno, Italy}
\affiliation[h]{ENEA, Fusion and Nuclear Safety Department, C.R. Frascati, Via E. Fermi 45, Frascati, I-00044 Rome, Italy}
\affiliation[i]{Physics Department, “Sapienza" University of Rome, P.le Aldo Moro 5, I-00185 Rome, Italy}
\affiliation[j]{Department of Mathematics, University of Wroclaw, plac Uniwersytecki 1, 50-137 Wrocław, Poland}
\affiliation[k]{Department of Statistics, Lund University, Box 117, SE-221 00 Lund, Sweden}

\emailAdd{maria.dainotti@nao.ac.jp}
\emailAdd{bdesimone@unisa.it}
\emailAdd{giovanni.montani@enea.it}
\emailAdd{malgorzata.bogdan20@gmail.com}

\abstract{\textbf{Abstract}\\
The standard cosmological model, the $\Lambda$CDM model, is the most suitable description for our universe. This framework can explain the accelerated expansion phase of the universe but still is not immune to open problems when it comes to the comparison with observations. One of the most critical issues is the so-called Hubble constant ($H_0$) tension, namely, the difference of about $5\sigma$ as an average between the value of $H_0$ estimated locally and the cosmological value measured from the Last Scattering Surface. The value of this tension changes from 4 to 6 $\sigma$ according to the data used. The current analysis explores the $H_0$ tension in the \textit{Pantheon} sample (PS) of SNe Ia. Through the division of the PS in 3 and 4 bins, the value of $H_0$ is estimated for each bin and all the values are fitted with a decreasing function of the redshift ($z$). Remarkably, $H_0$ undergoes a slow decreasing evolution with $z$, having an evolutionary coefficient compatible with zero up to $5.8\sigma$. If this trend is not caused by hidden astrophysical biases or $z$-selection effects, then the $f(R)$ modified theories of gravity represent a valid model for explaining such a trend.} 

\FullConference{%
  Multifrequency Behaviour of High Energy Cosmic Sources - XIV\\
  12-17 June 2023\\
  Mondello, Palermo, Italy
}

\begin{document}
\maketitle

\section{Introduction}
The $\Lambda$CDM model is based on the presence of Cold Dark Matter (CDM, not relativistic), the cosmological constant ($\Lambda$) that parametrizes the contribution of dark energy to the accelerated expansion of the universe, and $H_0$ that describes the expansion rate of the universe as observed today. The $\Lambda$CDM is the standard paradigm in modern cosmology but is not exempt from open issues. The most intriguing issue of $\Lambda$CDM is the \textit{$H_0$ tension}: this is the difference, ranging from 4 $\sigma$ up to 6 $\sigma$, between the local measurement $H_0$, obtained with the calibration of SNe Ia on the Cepheids, and the cosmological value of $H_0$ observed with the Cosmic Microwave Background (CMB) radiation at the redshift of the Last Scattering Surface, namely, $z=1100$. A wide range of solutions have been proposed to alleviate or to solve the $H_0$ tension, in particular, the interconnection between DM and Dark Energy (DE) to explain the Hubble tension, implying deviations from the $\Lambda$CDM model \citep{Bisnovatyi2023}. See \citep{DiValentino2021,Vagnozzi2023} for a more general review of the possible solutions for the $H_0$ tension. An important role in modern cosmology is played by SNe Ia, up to $z=2.26$ \citep{Rodney2016}. SNe Ia are among the best standard candles, given their uniform intrinsic luminosity.\\ 
In the current proceeding, we show the study of the $H_0$ tension through a bins approach applied to the PS of SNe Ia, spotlighting a slow decreasing evolution of the $H_0$ with redshift.

\section{The Pantheon sample analysis: unveiling new hints on the $H_0$ tension}\label{sec:H0tension}
We divided the PS \citep{Scolnic2018}, 1048 spectroscopically classified SNe Ia, into 3 and 4 bins ordered in redshift \citep{Dainotti2021SNe}. For each bin, we estimated the $H_0$ value through a Markov chain Monte Carlo simulation (MCMC), drawing the priors from a uniform distribution \\ ($60\,km/s/Mpc < H_0 < 80\,km/s/Mpc$) and fixing the SN absolute magnitude ($\mathcal{M}$) to the value $\mathcal{M}=-19.246$ so that $H_0=73.5\,km/s/Mpc$ in the lowest $z$ bins. Two cosmological models are employed: the $\Lambda$CDM and the $w_{0}w_{a}$CDM, where the Chevallier-Polarski-Linder parametrization \citep{CPL} is assumed for the equation of state parameter, $w(z)=w_{0}+(w_{a}*z/(1+z))$. $H_0$ is estimated in each bin through the minimization of the following $\chi^2$ model:

\begin{equation}
    \chi^2=\Delta \mu^{T} \mathcal{C}^{-1} \Delta \mu.
\end{equation}

\begin{figure}
    \centering
    \includegraphics[scale=0.12]{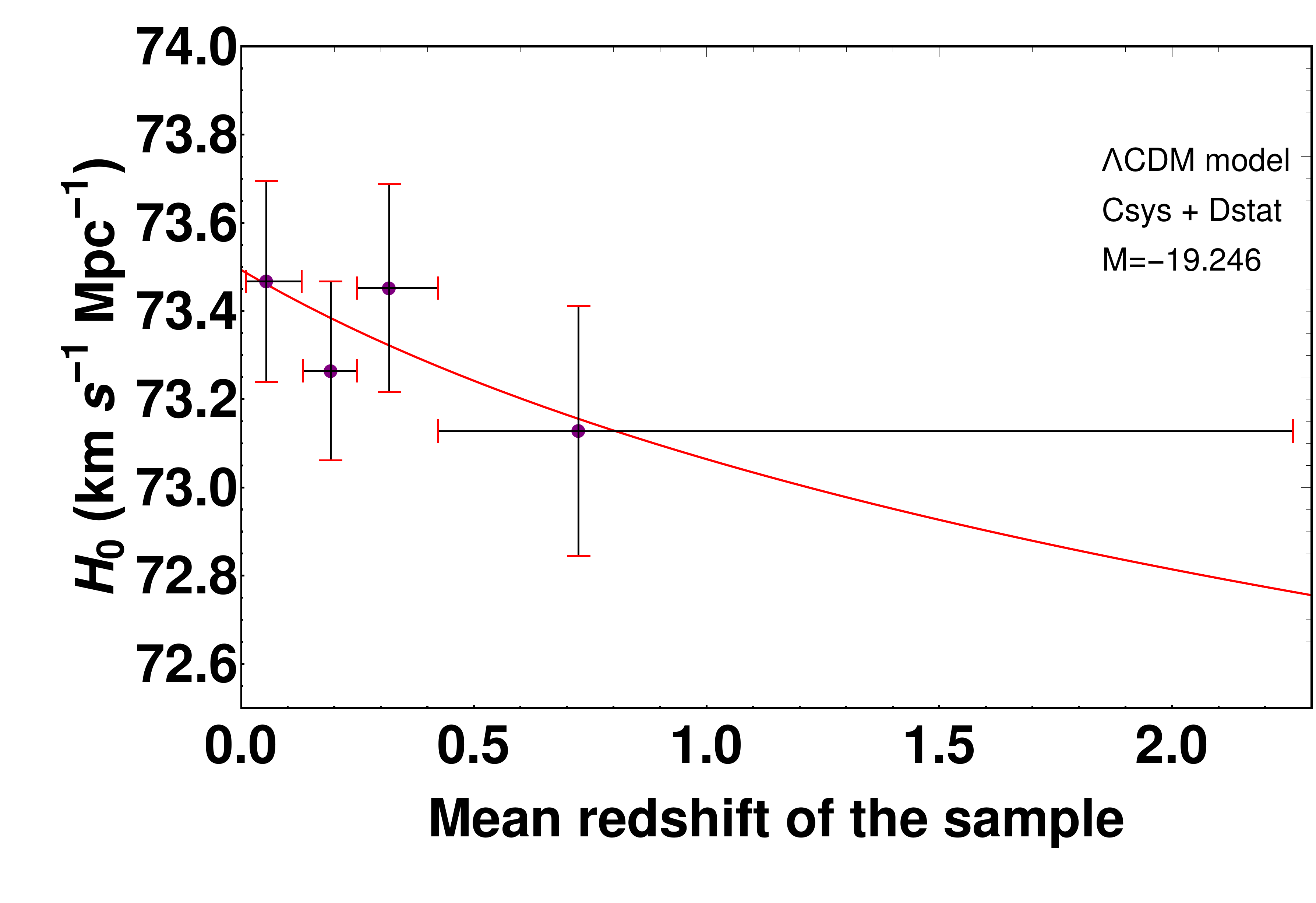}
    \includegraphics[scale=0.12]{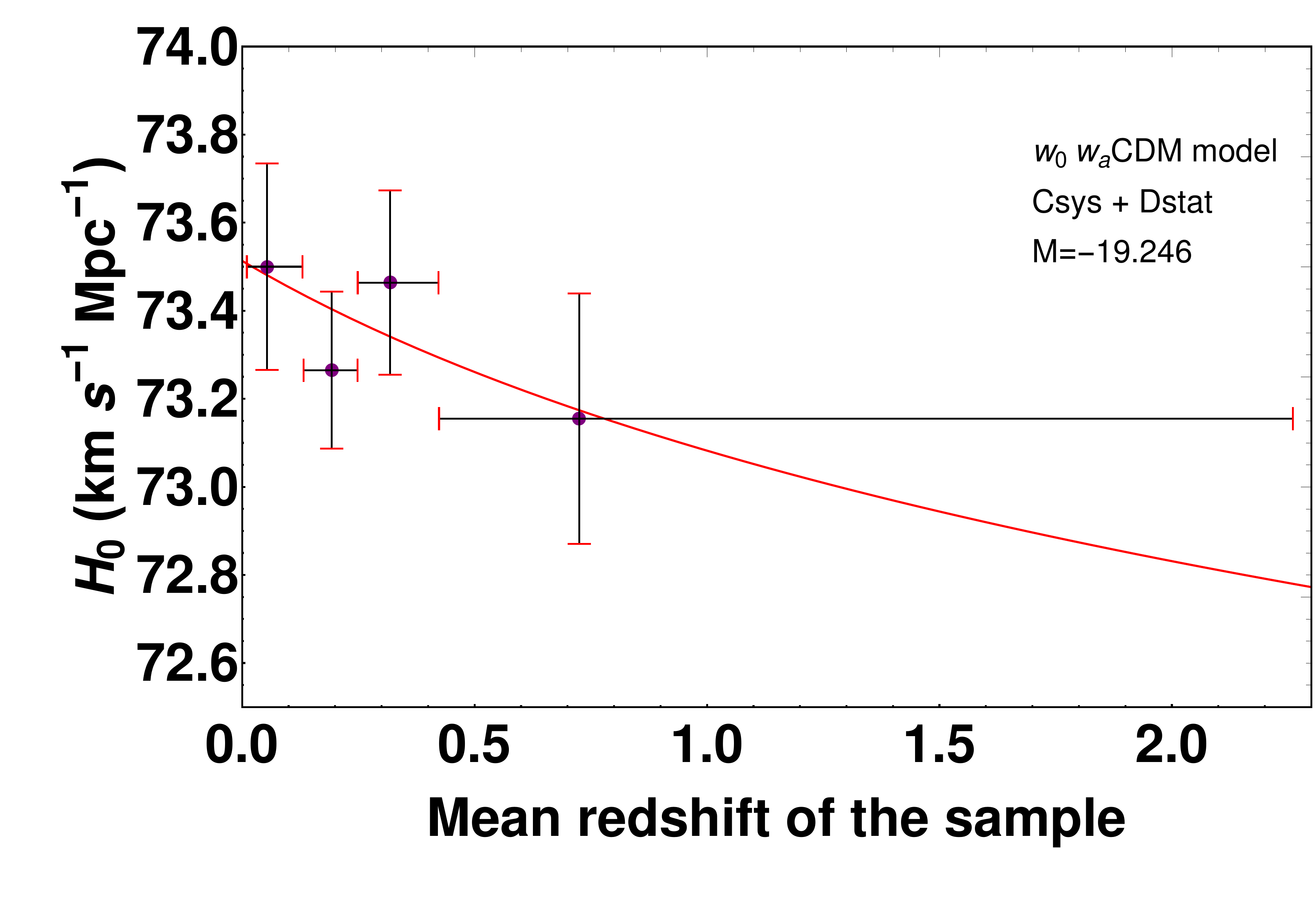}
    \caption{\textbf{Left panel.} The plotting of the $H_0$ values estimated in 4 bins together with the fitting $K(z)$ in the $\Lambda$CDM model. \textbf{Right panel.} The same plot in the case of $w_{0}w_{a}$CDM model. The figures are an excerpt of \citep{Dainotti2021SNe}.}
    \label{fig:H0(z)}
\end{figure}

Here $\Delta \mu$ is the difference between the theoretical distance moduli of SNe Ia and the observed one and $\mathcal{C}$ is the total covariance matrix of SNe Ia where both the contributions of systematic uncertainties ($C_{sys}$) and statistical uncertainties ($D_{stat}$) are included: $\mathcal{C}=C_{sys}+D_{stat}$.
The values of $H_0$ are then fitted with a function that decreases with $z$, described by the following equation:

\begin{equation}
    K(z)=\frac{H'_{0}}{(1+z)^q},
    \label{eq:H0z}
\end{equation}

where the fitting parameter $H'_{0}=H_0$ at $z=0$ and $q$ is the evolution parameter that describes the slope of the decreasing trend. We find that in the 3 bins division for the $\Lambda$CDM and $w_{0}w_{a}$CDM models, $q_{\Lambda}=0.009\pm0.004$ and $q_{w_{0}w_{a}}=0.008\pm0.004$ are compatible with the evolution scenario up to $2.0\,\sigma$. The fittings are reported in Figure \ref{fig:H0(z)}. 

We expanded the current analysis in \citep{Dainotti2022SNe} following the points reported below:

\begin{itemize}
\item Two parameters per time were left free to vary, $H_0,\Omega_{M}$ for the $\Lambda$CDM and $H_0,w_a$ for the $w_{0}w_{a}$CDM models, respectively;
\item The BAOs from \citep{BAOs} were added to the SNe Ia bins;
\item The PS is divided only into 3 bins to avoid the dominance of statistical fluctuations on the posterior distributions;
\item The priors have been drawn from the Gaussian distributions, expanded up to $2\,\sigma$, of the parameters $H_0,\Omega_{M},w_a$ where the mean values with $1\,\sigma$ uncertainties are $\Omega_{M}=0.298\pm0.022,H_0=70.393\pm1.079\,km/s/Mpc$, and $w_{a}=-0.129\pm0.026$, respectively;
\item The constraint $w(z)>-1$ is adopted to avoid the Phantom Dark Energy models;
\item The value of $\mathcal{M}=-19.35$ is fixed to the reference one reported in \citep{Scolnic2018}, implying $H_0=70\,km/s/Mpc$ locally, to avoid the statistical fluctuations given in the recalibration of SNe Ia when multiple parameters are left free to vary in the likelihood.
\end{itemize}

The decreasing trend is still present in the PS. In the analysis of \citep{Dainotti2021SNe,Dainotti2022SNe}, the compatibility with zero of the $q$ parameter ranges from $2.0\,\sigma$ in the $\Lambda$CDM model up $5.8\,\sigma$ in the $w_{0}w_{a}$CDM model with $q\propto 10^{-2}$. A possible explanation of this decreasing trend is the presence of unveiled astrophysical biases for SNe Ia light curve parameters, such as the stretch parameter following the discussion in \citep{Nicolas2021}. Alternatively, the interpretation of the $H_0$ decreasing trend via a fundamental physical effect can be performed via a metric $f(R)$-gravity in the so-called \textit{Jordan frame} \citep{Sotiriu,Odintsov}. The subtle question to be addressed concerns the typical feature of the most reliable proposals for interpreting the dark energy of the present universe \citep{HuSawicki}. 
In \citep{SchiavoneMontaniBombacigno}, it has been suggested to implement a metric $f(R)$-theory in the Jordan frame. The proposed scenario
relies on the idea, first discussed in \citep{Dainotti2021SNe}, that the effective behaviour $K(z)$ we observe from the data is due to the non-minimal coupling of the scalar mode present in the modified gravity formulation. This scalar field is associated with a potential term whose form is fixed by the considered $f(R)$ Lagrangian.
In solving the field equations for a flat Friedmann-Lemaitre-Robertson-Walker (FLRW, \citep{Robertson1935}) dynamics, we do not assign the form of the modified Lagrangian, but just the specific form of the scalar field $\phi(z)$, that account for the behaviour of $H_0 \propto (1+z)^{-q}$ we determined from data. Then two field equations (one is the generalized Friedmann equation, the other one is the result of variating the gravitational action with respect to $4\phi$) are then thought as fixing the two unknowns $H(z)$ and $V(z)$, $H(z)$ being the Hubble parameter and $V(z)$ being the searched potential term (i.e. the modified Lagrangian of gravity) which makes the dynamics a compatible system, respectively.
The dynamics are analyzed by including the matter density contribution of the present universe and generating the vacuum energy density as a contribution of the potential term, which is written as $V(\phi) = 2\chi \rho_{\Lambda} + U(\phi)$ (here $\chi$ is the Einstein constant). Thus, one of the two unknowns is specifically the function $U(z) \equiv U(\phi(z))$. We solved the equation analytically and numerically: in the former case, we make an approximation on the Friedmann equation, while the latter is treated in its exact form. The results obtained are consistent in the two cases. In particular, from the numerical profile of $U(z)$ and the adopted \textit{ansatz} for $\phi(z)$, we can reconstruct the
morphology of $U(\phi)$ (so that $V$ is a function of $\phi$, too, namely $V \equiv V(\phi)$). Hence, we can establish the form of the $f(R)$-Lagrangian. We study an analytical interpolation of the function $f$, which comes out at low $z$ values. The significant output of this analysis is that the modified gravity theory
appears to be a consistent proposal, and no tachyon mode emerges in its
dynamics. We can conclude that the observed smooth power-law $K(z)$ profile for SNe Ia is good evidence of the data analysis and can also be justified in a coherent proposal for modified gravity. The possibility of finding in the data set a valuable input for
possible new physics and from a consistently modified gravity proposal, a
justification for the observed behaviour has to be regarded as a robust feature of the proposed new scenario for the late universe.

\section{Summary and conclusions}
We have observed a consistent downward trend in the value of Hubble's constant ($H_0$) within the Pantheon Sample (PS). This trend, characterized by an evolutionary coefficient of approximately $10^{-2}$, holds statistical significance up to $5.8 \sigma$.
To explain this trend, we must consider the possibility that it's not a result of hidden biases in our selection process or the evolution of redshift ($z$) in the parameters of SNe Ia. If these factors can be ruled out, we may need to explore theories of modified gravity, specifically those falling under the $f(R)$ framework, as potential explanations for our current observations.
Since the $z$ range of SNe Ia is limited up to $z=2.26$, the $H_0$ tension calls for the use of high-$z$ probes as a cosmological tool, namely the Gamma-Ray Bursts (GRBs) that follow the fundamental plane relation \citep{DainottiPlatinum}.

\bigskip
\bigskip
\noindent {\bf DISCUSSION}

\bigskip
\noindent {\bf DEERAJ PASHAM:} It is clear that most of the discrimination power is beyond $z\sim 1.5$. How many SNe Type Ia do we have currently above $z\sim 1.5$? And how this will improve with upcoming all-sky-surveys like LSST and WFIRST?\\
{\bf Answer:} In the PS, only six SNe Ia are observed with a redshift $z \geq 1.5$, while in the recent PPS catalogue the SNe with $z \geq 1.5$ are seven. According to the simulation results reported in \citep{LSST}, the LSST is expected to observe up to 100 SNe Ia per year at redshift $z>1.2$. Considering the 10 years lifetime of the survey, this will correspond to an upper limit of 1000 SNe Ia at the end of the LSST. From the WFIRST SDT \citep{WFIRST} the total number of SNe Ia generated in the simulation is 2156 for redshift $0.8 \leq z \leq 1.7$, thus a large fraction of this will be observed at z $> 1.5$.


\begin{thebibliography}{99}

\bibitem{Bisnovatyi2023}
G. S. Bisnovatyi-Kogan \& A. M. Nikishin 2023, \emph{Eliminating the Hubble Tension in the Presence of the Interconnection between Dark Energy and Matter in the Modern Universe}, Astron. Rep., 67 2 115-124

\bibitem{DiValentino2021}
E. Di Valentino et al. 2021, \emph{Topical Review In the realm of the Hubble tension -  a review of solutions}, Class. Quantum Grav., 38 153001

\bibitem{Vagnozzi2023}
S. Vagnozzi 2023, \emph{Seven hints that early-time new physics alone is not sufficient to solve the Hubble tension}, Universe, 9(9) 393

\bibitem{Betoule2014}
M. Betoule et al. 2014, \emph{Improved cosmological constraints from a joint analysis of the SDSS-II and SNLS supernova samples}, A\&A, 568 A22 32

\bibitem{Rodney2016}
S. A. Rodney et al. 2015, \emph{Two SNe Ia at Redshift $\sim$ 2: Improved Classification and Redshift Determination with Medium-band Infrared Imaging}, ApJ, 150 5 156

















\bibitem{Scolnic2018}
D. M. Scolnic et al. 2018, \emph{The Complete Light-curve Sample of Spectroscopically Confirmed SNe Ia from Pan-STARRS1 and Cosmological Constraints from the Combined Pantheon Sample}, ApJ, 859 101

\bibitem{Dainotti2021SNe}
M. G. Dainotti et al. 2021, \emph{On the Hubble Constant Tension in the SNe Ia Pantheon Sample}, ApJ, 912 150

\bibitem{CPL}
Chevallier, M \& Polarski, D. 2001, \emph{Accelerating Universes with Scaling Dark Matter}, IJMPD, 10 213

\bibitem{Dainotti2022SNe}
M. G. Dainotti et al. 2022, \emph{On the Evolution of the Hubble Constant with the SNe Ia Pantheon Sample and Baryon Acoustic Oscillations: A Feasibility Study for GRB-Cosmology in 2030}, Galaxies, 10(1) 24

\bibitem{BAOs}
Sharov, G.S. \& Vasiliev, V.O. 2018, \emph{How predictions of cosmological models depend on Hubble parameter data sets}, MMG, 6

\bibitem{Nicolas2021}
N. Nicolas et al. 2021, \emph{Redshift evolution of the underlying type Ia supernova stretch distribution}, A\&A, 649 A74 

\bibitem{Sotiriu}
T. P. Sotiriu \& V. Faraoni 2010, \emph{f(R) Theories Of Gravity}, RMP, 82 451

\bibitem{Odintsov}
S. D. Odintsov et al. 2021, \emph{Analyzing the $H_0$ tension in $F(R)$ gravity models}, Nucl. Phys. B, 966, 115377

\bibitem{HuSawicki}
W. Hu \& I. Sawicki 2007, \emph{Models of $f(R)$ cosmic acceleration that evade solar system tests}, Phys. Rev. D 2007, 76, 064004

\bibitem{SchiavoneMontaniBombacigno}
T. Schiavone et al. 2023, \emph{$f(R)$ gravity in the Jordan Frame as a Paradigm for the Hubble Tension}, MNRAS, 522 1 L72-L77 

\bibitem{Robertson1935}
H. P. Robertson 1935, \emph{Kinematics and World-Structure}, ApJ, 82 284

\bibitem{DainottiPlatinum}
M. G. Dainotti et al. 2020, \emph{The X-Ray Fundamental Plane of the Platinum
Sample, the Kilonovae, and the SNe Ib/c Associated with GRBs}, ApJ, 904, 97

















\bibitem{WFIRST}
R. Hounsell et al. 2018, \emph{Simulations of the WFIRST Supernova Survey and Forecasts of Cosmological Constraints}, ApJ, 867 1 23 34

\bibitem{LSST}
LSST Science Collaboration: P. A. Abell et al. 2009, \emph{LSST Science Book, Version 2.0}, arXiv:0912.0201


\end{thebibliography}
\end{document}